\documentclass[conference, 10pt]{IEEEtran}
\IEEEoverridecommandlockouts
\usepackage{cite}
\usepackage{amsmath,amssymb,amsfonts}
\usepackage{algorithmic}
\usepackage{graphicx}
\usepackage{textcomp}
\usepackage{xcolor}
\usepackage{dsfont}
\usepackage{subcaption}
\usepackage{hyperref}
\def\BibTeX{{\rm B\kern-.05em{\sc i\kern-.025em b}\kern-.08em
    T\kern-.1667em\lower.7ex\hbox{E}\kern-.125emX}}

\begin{document}

\title{Adaptive Semantic Token Selection for AI-native Goal-oriented Communications
\thanks{This work has been supported by the SNS JU project 6G-GOALS under the EU’s Horizon program Grant Agreement No 101139232, by Sapienza grant RG123188B3EF6A80 (CENTS), and by European Union under the Italian National Recovery and Resilience Plan of NextGenerationEU, partnership on Telecommunications of the Future (PE00000001 - program RESTART).}
}

\author{
\IEEEauthorblockN{Alessio Devoto\IEEEauthorrefmark{1}\IEEEauthorrefmark{3}, Simone Petruzzi\IEEEauthorrefmark{1}, Jary Pomponi\IEEEauthorrefmark{2}\IEEEauthorrefmark{3}, Paolo Di Lorenzo\IEEEauthorrefmark{2}\IEEEauthorrefmark{3} and Simone Scardapane\IEEEauthorrefmark{2}\IEEEauthorrefmark{3}}
\IEEEauthorblockA{
\IEEEauthorrefmark{1}DIAG Department, Sapienza University of Rome, Via Ariosto 25, 00185, Rome, Italy}
\IEEEauthorblockA{
\IEEEauthorrefmark{3}Consorzio Nazionale Interuniversitario per le Telecomunicazioni, Viale G.P. Usberti, 181/A, 43124, Parma, Italy}
\IEEEauthorblockA{
\IEEEauthorrefmark{2}DIET Department, Sapienza University of Rome, Via Eudossiana 18, 00184, Rome, Italy\\ Emails: \emph{\{firstname.lastname\}@uniroma1.it}}
}

\maketitle

\begin{figure*}[ht!]
\centering
  \includegraphics[width=0.9\textwidth]{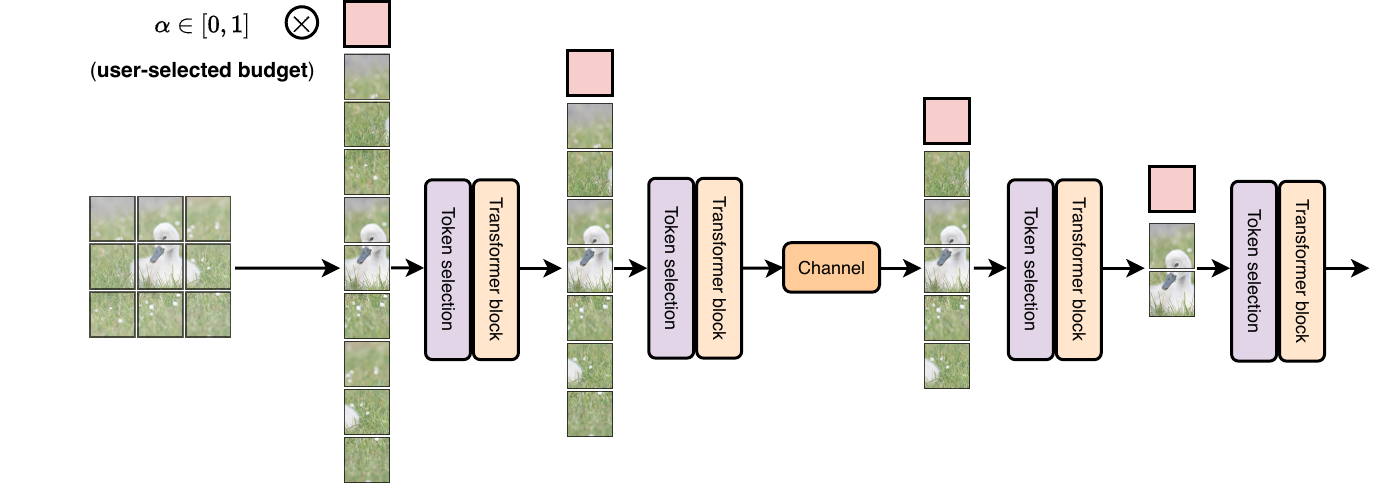}
  \caption{Overview of the proposed method. Each transformer block is preceded by a trainable token selection block, which leverages a user-provided runtime budget to discard tokens. Different budgets result in different model behaviours. During training, the budget is randomly selected, at inference the budget is selected by the user.} \label{fig:general-pipeline}
\end{figure*}

\begin{abstract}
In this paper, we propose a novel design for AI-native goal-oriented communications, exploiting transformer neural networks under dynamic inference constraints on bandwidth and computation. Transformers have become the standard architecture for pre-training large-scale vision and text models, and preliminary results have shown promising performance also in deep joint source–channel coding (JSCC). Here, we consider a dynamic model where communication happens over a channel with variable latency and bandwidth constraints.  Leveraging recent works on conditional computation, we exploit the structure of the transformer blocks and the multihead attention operator to design a trainable \textit{semantic token selection} mechanism that learns to select relevant tokens (e.g., image patches) from the input signal. This is done dynamically, on a per-input basis, with a rate that can be chosen as an additional input by the user. We show that our model improves over state-of-the-art token selection mechanisms, exhibiting high accuracy for a wide range of latency and bandwidth constraints, without the need for deploying multiple architectures tailored to each constraint. Last, but not least, the proposed token selection mechanism helps extract powerful semantics that are easy to understand and explain, paving the way for \textit{interpretable-by-design} models for the next generation of AI-native communication systems.
\end{abstract}

\vspace{.1cm}

\section{Introduction}

Recently, interest in semantic and goal-oriented communication has grown, identifying this new paradigm as a keystone for exploiting the semantic meaning of data, and enabling effective task execution among distributed network nodes \cite{strinati20216g,strinati2024goal}. Among possible approaches, joint source-channel coding (JSCC) exploiting neural networks for simultaneous coding and transmission over wireless channels (referred to as \textit{deep} JSSC schemes \cite{xu2023deep}) have become popular. Deep JSCC methods leverage the powerful approximation capabilities of modern neural networks (NNs), such as convolutional NNs (CNNs) \cite{bourtsoulatze2019deep} and transformers \cite{xie2021deep}, to optimize coding and transmission for specific tasks, by incorporating the transmission channel as a non-trainable layer inside the network and optimizing the complete system end-to-end. These methods can be shown to outperform standard coding schemes in semantic communication scenarios, with a graceful degradation in performance that avoids steep declines once the signal-to-noise (SNR) ratio falls below information-theoretic thresholds \cite{bourtsoulatze2019deep}.

In the simplest case, one neural network is trained to maximize the performance relative to a given task (e.g., reconstruction) given a model of the channel's condition. This provides a \textit{fixed} trade-off in terms of computation, bandwidth, and accuracy for that channel. Practically, we may be interested in \textit{adaptive} models, able to select the appropriate compression rate and computational budget based on the input itself or on the channel condition, either in terms of latency (i.e., proportional to the number of operations that are executed by the model) or in terms of bandwidth (i.e., proportional to the size of the message which is transmitted over the communication channel) \cite{xu2023deep}. To achieve this, one can train separate models with different trade-offs, or embed the adaptivity inside the model itself. 
Recently, transformer optimization methods that initially emerged outside of the communication domain, such as quantization \cite{courbariaux2014training, wu2020integer,dettmers2022llm}, distillation \cite{hinton2015distilling,aguilar2020knowledge}, sparsification \cite{lecun1989optimal,hoefler2021sparsity}, and lately dynamic networks \cite{wang2021not, avit, wojcik2023adaptive}, have been successfully adopted in the JSCC scenario to enforce latency or bandwidth constraints.
In the bandwidth case, several authors have proposed trainable techniques for designing adaptive deep JSCC schemes, such as partitioning the latent space of the model \cite{xu2023deep}, masking non-relevant channels in a CNN \cite{yang2022deep}, or projecting each token of a transformer to variable-length vectors \cite{dai2022nonlinear}. 
For example, in \cite{wang2023adaptive}, the authors proposed an adaptive source allocation paradigm based on quantization, which improves transmission reliability, while in \cite{zhou2021semantic}, a Universal Transformer is trained to achieve flexible communication, enabling better performance under various channel conditions. 

By contrast, in this paper we propose a principled design for transformer-based JSCC algorithms, able to satisfy both bandwidth and computational (latency) constraints in a unified way. While previous work has considered compressing information at variable rates across the embedding dimension, we show promising results in the design of a model able to select only the most relevant tokens to transmit. This selection is \textit{trainable} and \textit{dynamic}, in the sense that the amount of tokens that are selected or discarded can be chosen by the user with an additional \textit{budget} parameter at inference time, while during training we optimize the model for every possible choice of budget. We provide several numerical results showing how the proposed strategy outperforms alternative token selection and compression baselines in terms of accuracy, communication efficiency, and interpretability.

\section{System Model}

The proposed deep JSCC model for goal-oriented semantic communications is depicted in Fig. \ref{fig:general-pipeline}. We first describe the standard deep JSCC model in Sec. \ref{sec:channel_model}, and then we discuss our proposed adaptive token selection mechanism in Sec. \ref{sec:method}.

\subsection{Deep JSCC model}
\label{sec:channel_model}

We consider a deep JSCC model where the entire communication pipeline is described by a neural network $f(x) = (E \odot C \odot D)(x)$, where $E$ is a transformer that maps the inputs to the channel's symbols, $C$ is a set of (non-trainable) layers that simulate the channel, and $D$ is a second transformer model that maps the received symbols to the task under consideration.

The encoder first maps the input $x$ to a set of tokens $\mathbf{x} \in \mathbb{R}^{n \times d}$, where $n$ is the number of tokens (e.g., the number of image patches in a vision transformer), and $d$ is the embedding dimension. Then, $\mathbf{x}$ is processed by a series of $L_e$ transformer blocks to output a second matrix $\mathbf{h}$, typically of the same shape, where each token represents a symbol to be transmitted. For the channel $C$ we follow a standard setup \cite{xu2023deep} and we implement it as a layer adding noise to the symbols with a pre-specified power. Specifically, we add noise based on a specific Signal-to-Noise Ratio (SNR), in which case the received signal $\mathbf{h}^\prime$ at the decoder can be expressed as:
\begin{equation} 
\mathbf{h}^\prime = \mathbf{h} + \mathbf{n} 
\end{equation}
where $\mathbf{n}$ is drawn from a zero-mean complex Gaussian distribution with variance $\sigma^2$, such that the SNR is given by:
\begin{equation}
\text{SNR}_{\text{db}}(x) = 10 \log_{10}\left(\frac{{\lVert \mathbf{h} \rVert^2}}{{\lVert \mathbf{n} \rVert^2}}\right)   
\end{equation}
We also experiment the robustness of our model to possible packet drops, where the received signal $\mathbf{h}^\prime$ is obtained by randomly removing rows from the transmitted signal $\mathbf{x}$ with a dropping probability $p_d$. By changing the SNR for the first setup and $p_d$ for the second setup we can control the additive noise intensity. The decoder $\hat{y} = D(\mathbf{h}^\prime)$ transforms the received tokens with $L_d$ additional transformer blocks to output a prediction $\hat{y}$ which depends on the communication goal. The model $f$ is trained end-to-end by minimizing some loss $\mathcal{L}(\hat{y}, y)$, e.g., a cross-entropy loss for classification.

\subsection{Adaptive semantic token selection}
\label{sec:method}

We extend the deep JSCC model from Section \ref{sec:channel_model} to consider variable constraints at inference, either in terms of the channel's bandwidth (how many symbols we transmit) or compute power (how many operations the model is allowed to perform). We frame both into a shared formulation and we show in Sec. \ref{sec:penalties} specific customizations for the two cases. Denote by $T$ the cost (time or bandwidth) required by the original network $f$. We augment the network with an additional parameter $\alpha \in \left[0,1\right]$, that we call the \textit{budget}, such that running $f(x,\alpha)$ should be done in less than $\alpha T$:
\begin{align}
    \min_f & \;\; \mathbb{E}_{(x,y)} \left[ \mathcal{L}(f(x, \alpha), y) \right] \nonumber\\
    \text{s.t.} & \;\; T(x) \le \alpha T \;\; \forall \alpha \in \left[0,1\right]
    \label{eq:formulation}
\end{align}
where $T(x)$ measures the cost of running $f$ on $x$. For a standard model, the cost $T(x) = T$ is fixed and \eqref{eq:formulation} cannot be satisfied. To this end, we further augment $E$ and $D$ with additional trainable layers to reduce the cost adaptively. In particular, as illustrated in Fig. \ref{fig:general-pipeline}, we augment each transformer block with trainable layers that select a subset of tokens to be removed, progressively over the entire architecture.

\begin{figure}[t]
\centering
  \includegraphics[width=0.35\textwidth]{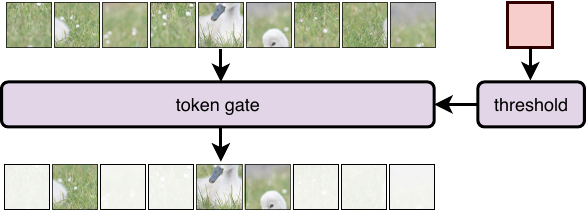}
  \caption{Detail of the token selection module. The module learns a threshold $\gamma_k$ from the budget token and a halting score $s_i$ for other tokens. Tokens are discarded if their score is lower than the learned threshold.} \label{fig:zoom}
\end{figure}

First, we represent the budget by a trainable token which is multiplied by $\alpha$ and appended to the input tokens. Our adaptive selection module comprises two linear layers that we call \textit{threshold selector} and \textit{token gate}. As depicted in Fig. \ref{fig:zoom}, at a generic layer $k$ the budget token goes through a linear layer to compute a gate threshold $\gamma_k \in (0,1)$. Then, the token gate outputs, for each token $i$, a single scalar $g_i$, and we compute the halting score $s_i \in (0,1)$ as:
\begin{equation}
s_i = \sigma( \delta \cdot g_i + \beta)  
\end{equation}
where $\sigma$ is the logistic sigmoid function and  $\delta, \beta$ are terms to control the sigmoid slope and bias. The token is then dropped if $ s_i < \gamma_k$, i.e., it is removed from all subsequent computations. To handle the constraint in \eqref{eq:formulation}, we recast it as an unconstrained optimization problem as follows:
\begin{equation}
\min_f \;\; \mathbb{E}_{\alpha} \left[ \mathbb{E}_{(x,y)} \left[ \mathcal{L}(f(x, \alpha), y) + \lambda (T(x) - \alpha T)^2 \right] \right] 
\label{eq:formulation_unconstrained}
\end{equation}
where $\lambda$ is an additional regularization parameter. In practice, we approximate the first expectation by randomly sampling $\alpha$ in the uniform range $[0,1]$ during training for each mini-batch. As long as the cost $T(x)$ is differentiable, \eqref{eq:formulation_unconstrained} can be solved with standard gradient descent.

\subsection{Designing the cost penalties}
\label{sec:penalties}

For each layer, the sparsity of the input sequence depends on how many tokens have not been discarded up to that point, i.e. for how many tokens $i$ we have $s_i \geq \gamma_k$. Hence, we define the sparsity of an input $x$ at layer $k$ as 
\begin{equation}
S_k = \frac{1}{n} \sum_{i=1}^{n} \max(s_i - \gamma_k, 0)
\end{equation}
where $n$ is the initial number of tokens. In low-latency applications, we are interested in the global amount of computational resources spent by the model, so we average the sparsity across all layers and we define the cost as:
\begin{equation}
    T(x) = \frac{1}{L_e + L_d} \sum_{k=1}^{L_d + L_e} S_k
    \label{eq:global_loss}
\end{equation}

\noindent where the index $k$ runs over all layers $k$ across both encoder and decoder. We call \eqref{eq:global_loss} a \textit{global} penalty. This formulation gives the model more freedom, as the budget can be distributed across the layers as needed. In the case we have a bandwidth constraint, we define the cost as the sparsity at the output layer of the encoder, which we denote generically by $e$:
\begin{equation}
    T(x) =  S_e
    \label{eq:local_loss}
\end{equation}
We call this a \textit{local} penalty. This second choice results in a bottleneck right before the channel, so that the model is forced to discard sufficient tokens before transmitting the symbols. In this case, the decoder is not affected by the regularization term and does not have token selection layers.

Once the model is trained, we can control the budget by manually setting the parameter $\alpha \in [0, 1]$, as in Fig. \ref{fig:general-pipeline}. If the model was trained using the global penalty term (Eq. \eqref{eq:global_loss}), the budget acts as a soft bound for the FLOPs, and, as a consequence, on the latency. Otherwise, when the local penalty (Eq. \eqref{eq:local_loss}) is used, the budget acts as a constraint on the channel's bandwidth.

\section{Numerical Results}

We validate our method on the Imagenette dataset for image classification. Following \cite{avit}, we use the pre-trained Data Efficient Vision Transformer \cite{deit} (DeiT) as starting model. We run two separate sets of experiments to validate the two penalties from Sec. \ref{sec:penalties}, and we leave training them simultaneously as future work. In the first set of experiments, we use only the global penalty in \eqref{eq:global_loss} to reduce the inference latency of the complete model. In the second set of experiments, we force token selection at the output of the encoder using instead the local penalty in \eqref{eq:local_loss}. 

\begin{figure}[t]
\centering
  \includegraphics[width=0.4\textwidth]{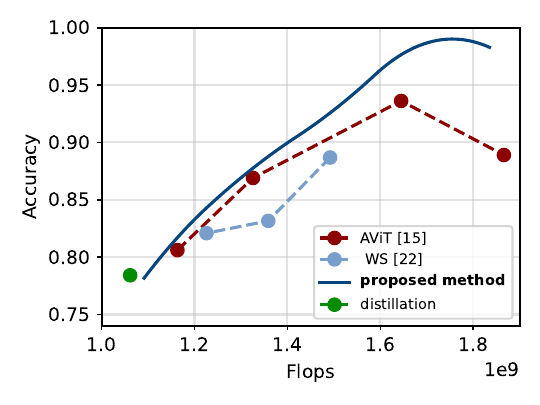}
  \caption{Accuracy-efficiency trade-off (in FLOPs) for the proposed method compared to other baselines, in noiseless setup. Unlike other methods that require fine-tuning for each budget (separate markers), our method delivers a \textit{single} model for all possible budgets, represented as a continuous line.} 
  \label{fig:imagenette-acc-flops}
\end{figure}

We compare our results with one state-of-the-art conditional computation method (AViT \cite{avit}) as well as two standard techniques which do not fall in the category of conditional computation: distillation \cite{hinton2015distilling,aguilar2020knowledge} and Weight Subcloning \cite{weight_cloning}. Both techniques provide a fine-tuned model with a fixed inference budget, and we train separate models on a range of different budgets for comparison.  For AViT \cite{avit}, we finetune the model for token depths of 4,7,9, and 11. For Weight Subcloning \cite{weight_cloning}, we remove either one, two or three central layers. Finally, for distillation, we use a hidden features size of 96 and 10 layers.

We measure the impact of token selection on inference latency using FLOPs (FLoating Point Operations). We show the results of applying our method with the global penalty from Eq. \eqref{eq:global_loss} in Fig. \ref{fig:imagenette-acc-flops}, where we run experiments in a noiseless setup. We highlight again that our method trains a \emph{single} model (shown as a continuous line), while the baselines always imply a separate fine-tuning for each budget (shown as dashed lines). 

\begin{figure}[t]
\centering
  \includegraphics[width=0.4\textwidth]{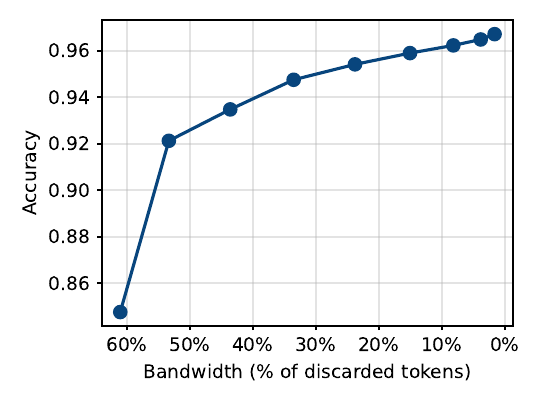}
  \caption{Accuracy of the proposed method when trained with the local penalty. We impose a budget constraint on the number of discarded tokens (bandwidth), represented on the x-axis. We do not show the baselines here as they do not allow for adaptive badnwidth selection.} \label{fig:imagenette-bandwidth-acc}
\end{figure}

The proposed model achieves competitive results across a wide range of budget choices. In Fig. \ref{fig:imagenette-bandwidth-acc} we show the results of token selection trained with the local penalty from Eq. \eqref{eq:local_loss}. In this case, we use the percentage of discarded tokens per each image as a proxy for the bandwidth of the channel, and we don not show the baselines as they do not allow for adaptive bandwidth selection.

\subsection{Noisy channel}
\begin{figure}[t]
\centering
  \includegraphics[width=0.4\textwidth]{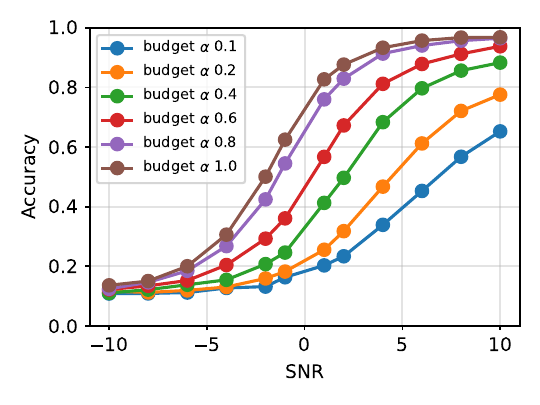}
  \caption{Accuracy of the proposed method across different SNRs. Each line represents a budget, imposed at inference via the budget token. If the current channel condition is known, selecting the right budget can save resources.} \label{fig:imagenette-analog-noise}
\end{figure}

\begin{figure}[t]
\centering
  \includegraphics[width=0.4\textwidth]{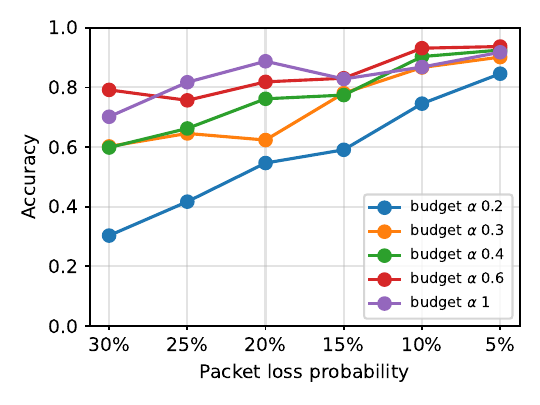}
  \caption{Accuracy of the proposed method for different probabilities of dropping a token. Each line represents a budget, imposed at inference via the budget token.} \label{fig:imagenette-digital-noise}
\end{figure}

We test a noisy channel between the two central blocks of the model as described in Section \ref{sec:channel_model}. For each type of noise, we show the model's behaviour for different budgets, imposed at runtime via the $\alpha$ parameter. For Gaussian noise, we control the noise severity by increasing the SNR, whereas for packet dropping, we assume the probability of token dropping as a proxy for the loss of packets.  In both cases, we average the experiments over 5 different runs. 
We show the results in Fig. \ref{fig:imagenette-analog-noise} and Fig. \ref{fig:imagenette-digital-noise}, respectively. As we can see, the model proves to be robust to channel impairments in both scenarios. In the case of packet dropping, we observe that for lower packet loss probabilities (up to 10\%) we achieve a surprisingly small accuracy drop when decreasing the budget.

\subsection{Interpretability of token selection}

\begin{figure}[h]
    \centering
    \begin{subfigure}[b]{0.2\textwidth}

        \centering
        \includegraphics[width=\textwidth]{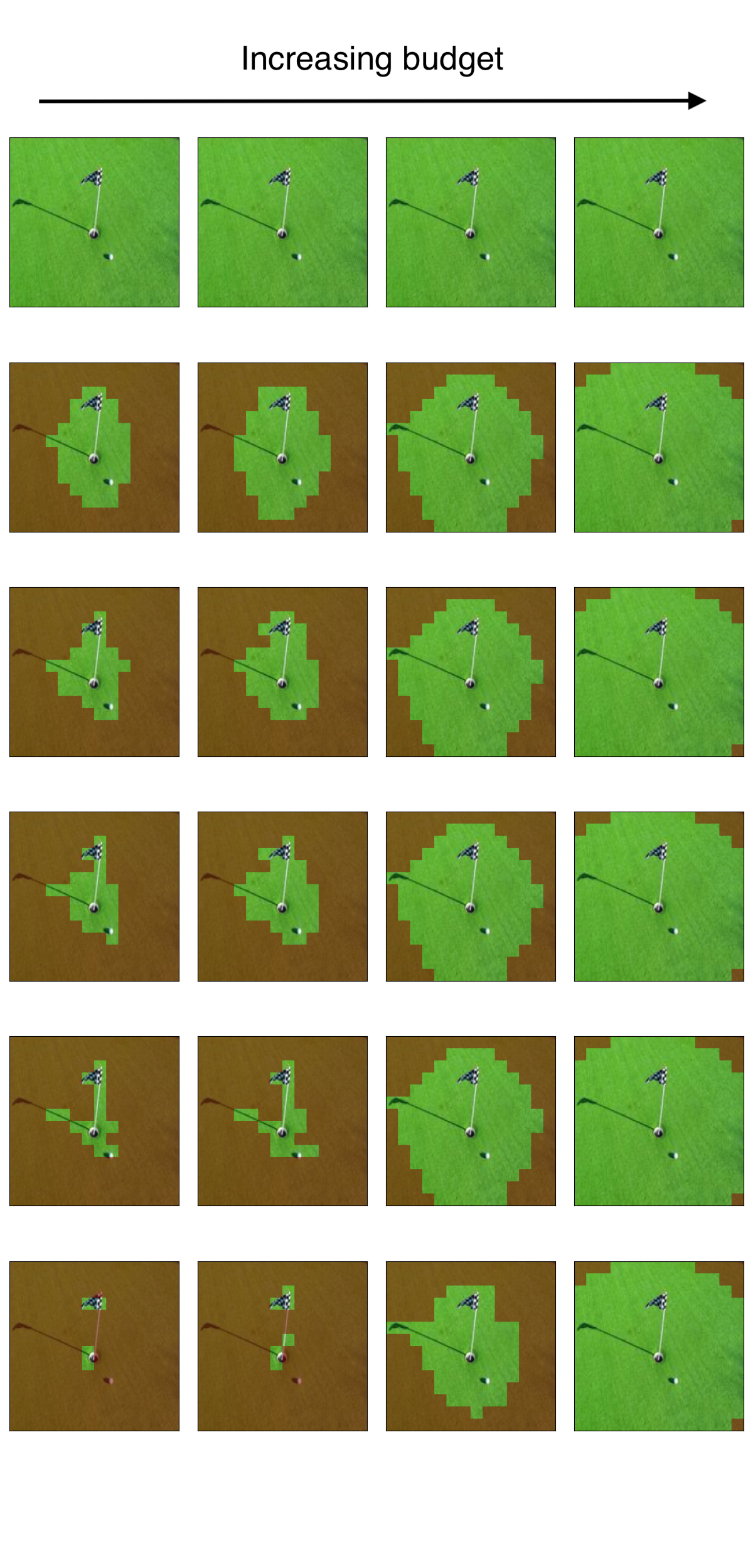}
    \end{subfigure} \hspace{0.5em}
    \begin{subfigure}[b]{0.2\textwidth}
        \centering
        \includegraphics[width=\textwidth]{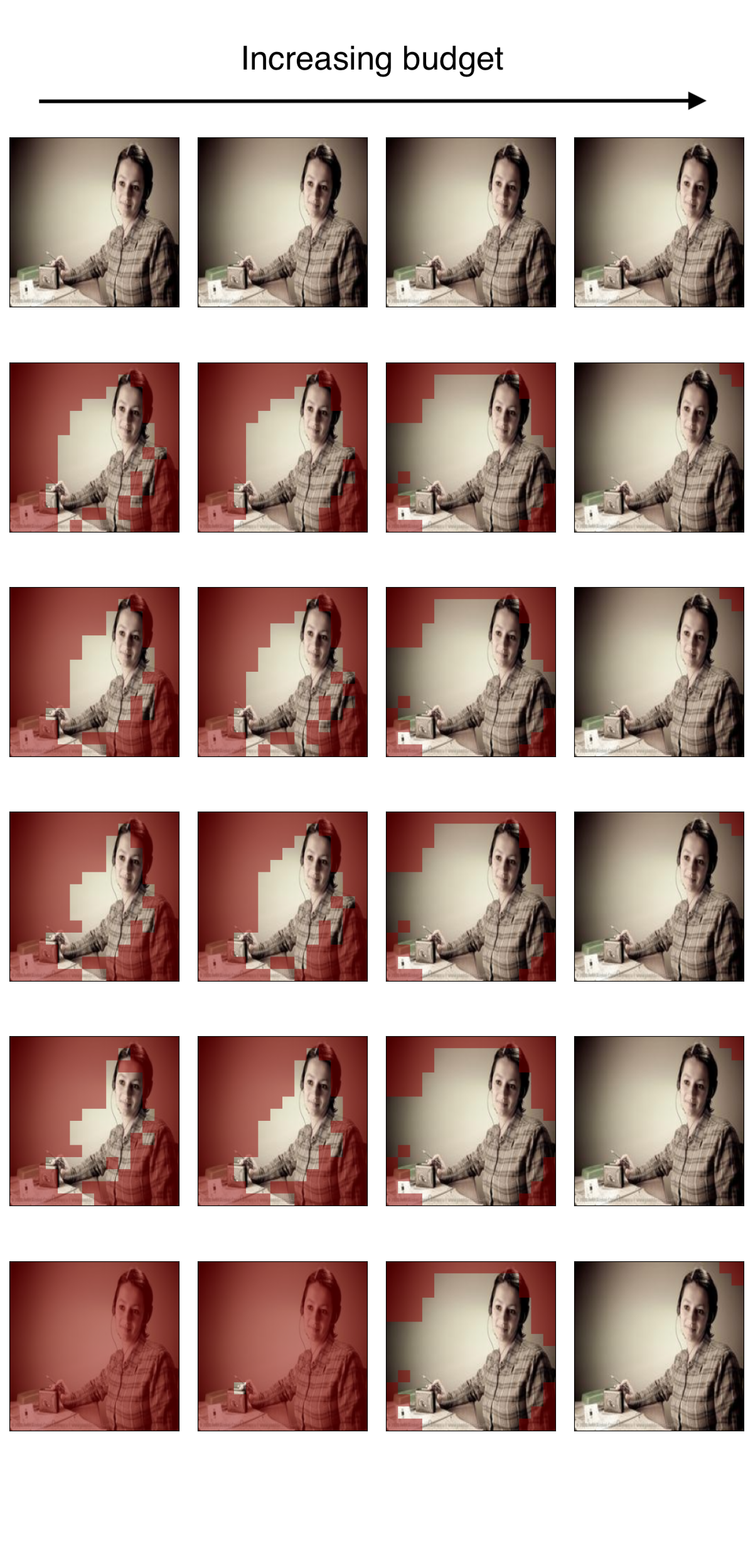}
    \end{subfigure}
\caption{Tokens discarded at layers 1, 3, 5, 7, 9 and 11 of the model for different budgets (0.3, 0.5, 0.7, 1), using the \textbf{global} penalty. Each row represents a transformer layer, and each column a different budget. The model seems to keep tokens that align with the semantic meaning of images. For easier samples, the model learns to skip the last layer.} \label{fig:discarded-tokens-global}
\end{figure}

\begin{figure}[h]
    \centering
    \begin{subfigure}[b]{0.2\textwidth}
        \centering
        \includegraphics[width=\textwidth]{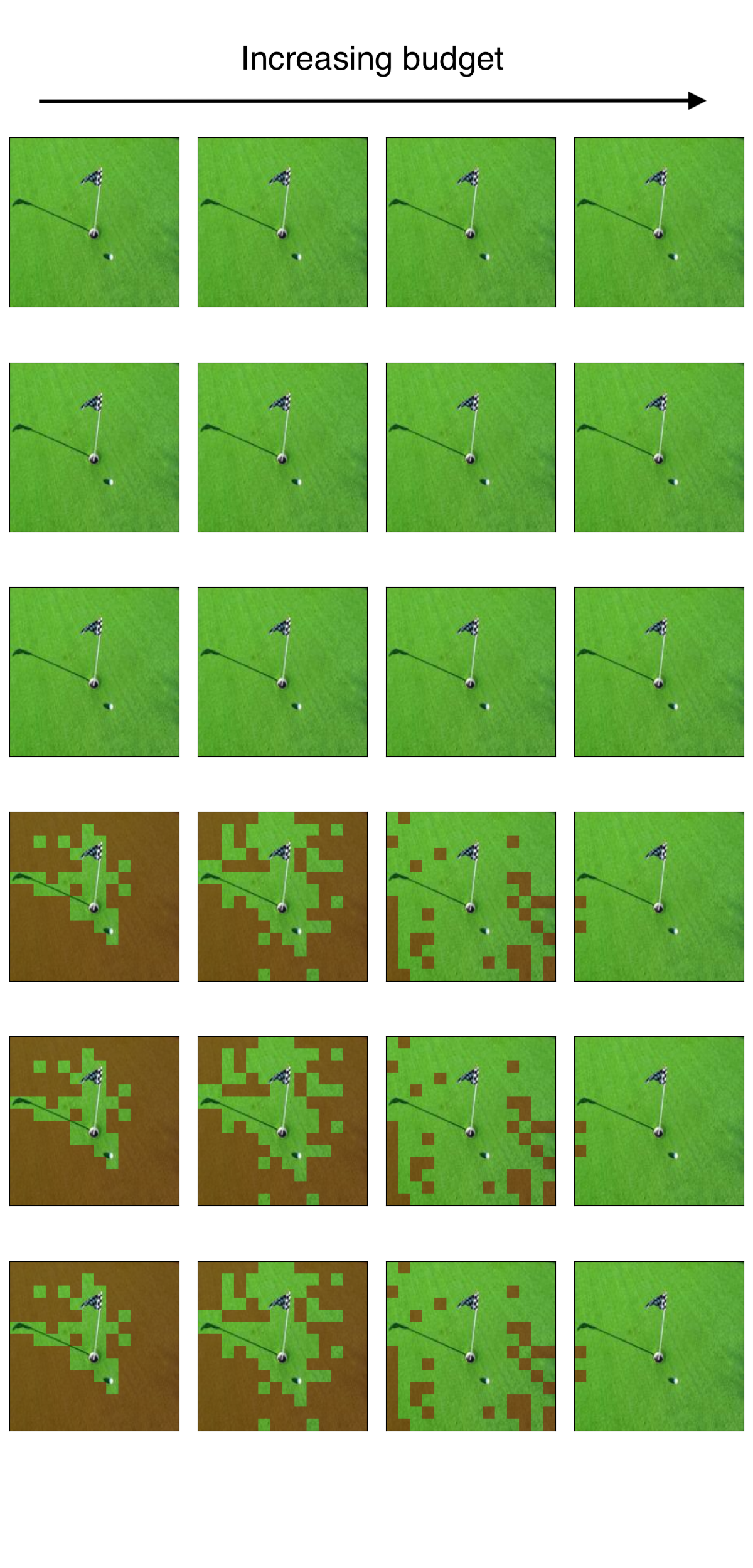}
    \end{subfigure}  \hspace{0.5em}
    \begin{subfigure}[b]{0.2\textwidth}
        \centering
        \includegraphics[width=\textwidth]{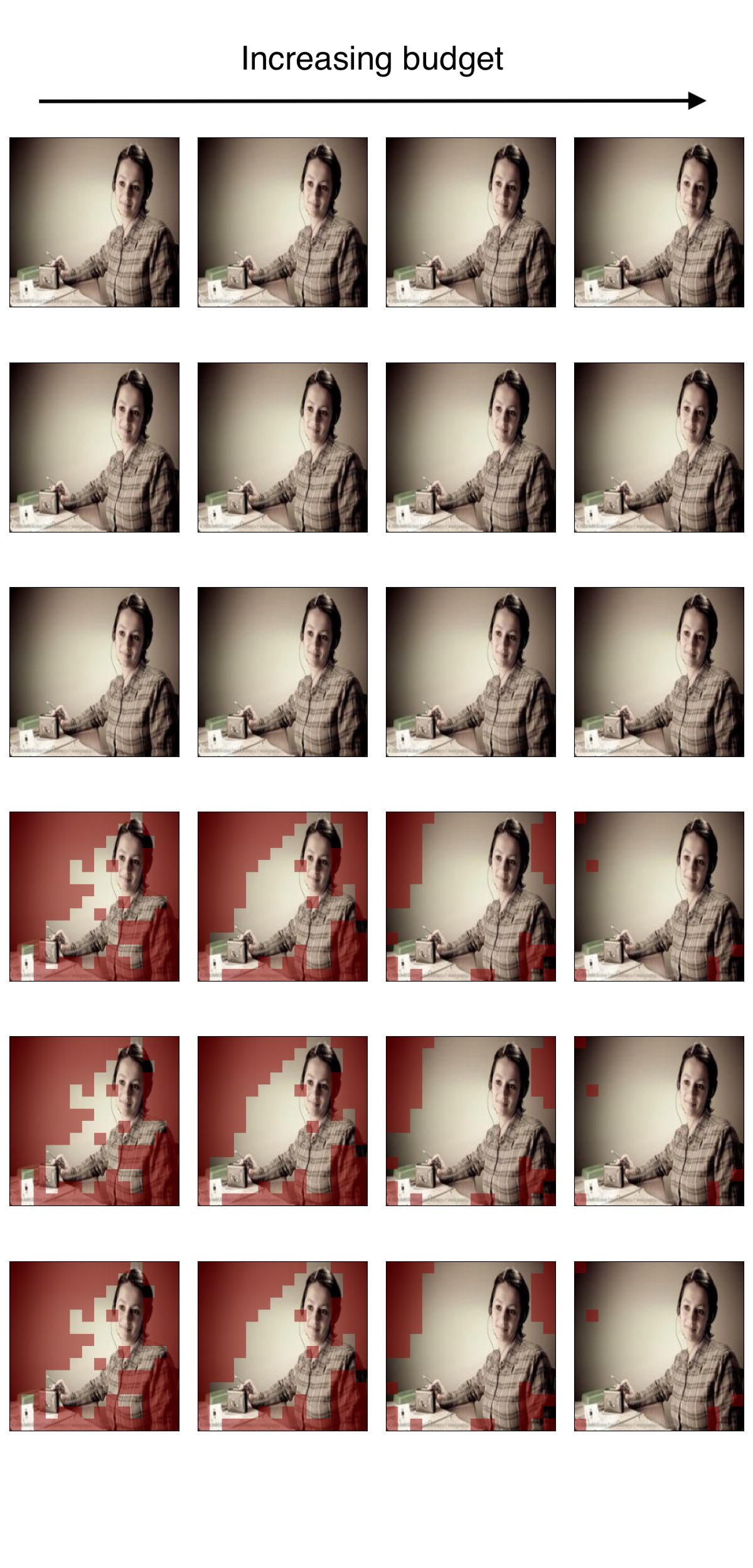}
    \end{subfigure}
\caption{Tokens discarded at layers 1, 3, 5, 7, 9 and 11 of the model for different budgets (0.3, 0.5, 0.7, 1), using the \textbf{local} penalty. Each row represents a transformer layer, and each column a different budget. The model learns to use all the tokens and discard them only before the channel.} \label{fig:discarded-tokens-local}
\end{figure}

A key feature of our approach is the possibility to visualize which tokens are discarded by the model at each layer when using global (Fig. \ref{fig:discarded-tokens-global}) and local (Fig. \ref{fig:discarded-tokens-local}) penalties. Indeed, as we will see, the masks often align with the semantic meaning of tokens. This is probably due to the fact that the model, especially when forced to use low budgets, learns to keep the most semantically meaningful tokens, i.e. the ones carrying information relevant for the specific task at hand. 
Interestingly, we observe that for some "easy" samples the model skips the last layer, resulting in a behaviour similar to an early exit mechanism.
When we use the local penalty from Eq. \eqref{eq:local_loss}, the model is forced to discard tokens locally, right before the channel. The model behavior aligns surprisingly well with this constraint. The initial layers process all the tokens, retaining the maximum information, and tokens are only discarded before transmission. This behaviour is clear in Fig. \ref{fig:discarded-tokens-local}. 

\section{Conclusions and future Work}
In this work, we have proposed a novel design for goal-oriented semantic communications, hinging on a transformer-based JSCC scheme that incorporates a semantic token selection mechanism to adaptively handle time and bandwidth constraints. Our model outperforms current token selection mechanisms, offering high accuracy across diverse constraints and robustness to channel impairments. Additionally, the proposed semantic token selection inherently supports the development of interpretable AI-native communication systems. In future work, we plan to experiment with different modalities, as we conjecture that for other tasks the model might learn to compress information in an even smaller amount of tokens. Furthermore, we plan to extend the method by adding empty trainable tokens which could serve as a memory to store and retrieve information after transmission.


\begin{thebibliography}{10}
\providecommand{\url}[1]{#1}
\csname url@samestyle\endcsname
\providecommand{\newblock}{\relax}
\providecommand{\bibinfo}[2]{#2}
\providecommand{\BIBentrySTDinterwordspacing}{\spaceskip=0pt\relax}
\providecommand{\BIBentryALTinterwordstretchfactor}{4}
\providecommand{\BIBentryALTinterwordspacing}{\spaceskip=\fontdimen2\font plus
\BIBentryALTinterwordstretchfactor\fontdimen3\font minus \fontdimen4\font\relax}
\providecommand{\BIBforeignlanguage}[2]{{%
\expandafter\ifx\csname l@#1\endcsname\relax
\typeout{** WARNING: IEEEtran.bst: No hyphenation pattern has been}%
\typeout{** loaded for the language `#1'. Using the pattern for}%
\typeout{** the default language instead.}%
\else
\language=\csname l@#1\endcsname
\fi
#2}}
\providecommand{\BIBdecl}{\relax}
\BIBdecl

\bibitem{strinati20216g}
E.~C. Strinati and S.~Barbarossa, ``{6G} networks: Beyond {S}hannon towards semantic and goal-oriented communications,'' \emph{Computer Networks}, vol. 190, p. 107930, 2021.

\bibitem{strinati2024goal}
E.~C. Strinati, P.~Di~Lorenzo, V.~Sciancalepore, A.~Aijaz, M.~Kountouris, D.~G{\"u}nd{\"u}z, P.~Popovski, M.~Sana, P.~A. Stavrou, B.~Soret \emph{et~al.}, ``Goal-oriented and semantic communication in {6G} {AI}-native networks: The {6G-GOALS} approach,'' \emph{arXiv preprint arXiv:2402.07573}, 2024.

\bibitem{xu2023deep}
J.~Xu, T.-Y. Tung, B.~Ai, W.~Chen, Y.~Sun, and D.~D. G{\"u}nd{\"u}z, ``Deep joint source-channel coding for semantic communications,'' \emph{IEEE Communications Magazine}, vol.~61, no.~11, pp. 42--48, 2023.

\bibitem{bourtsoulatze2019deep}
E.~Bourtsoulatze, D.~B. Kurka, and D.~G{\"u}nd{\"u}z, ``Deep joint source-channel coding for wireless image transmission,'' \emph{IEEE Trans. on Cognitive Communications and Netw.}, vol.~5, no.~3, pp. 567--579, 2019.

\bibitem{xie2021deep}
H.~Xie, Z.~Qin, G.~Y. Li, and B.-H. Juang, ``Deep learning enabled semantic communication systems,'' \emph{IEEE Transactions on Signal Processing}, vol.~69, pp. 2663--2675, 2021.

\bibitem{courbariaux2014training}
M.~Courbariaux, Y.~Bengio, and J.-P. David, ``Training deep neural networks with low precision multiplications,'' \emph{arXiv preprint arXiv:1412.7024}, 2014.

\bibitem{wu2020integer}
H.~Wu, P.~Judd, X.~Zhang, M.~Isaev, and P.~Micikevicius, ``Integer quantization for deep learning inference: Principles and empirical evaluation,'' \emph{arXiv preprint arXiv:2004.09602}, 2020.

\bibitem{dettmers2022llm}
T.~Dettmers, M.~Lewis, Y.~Belkada, and L.~Zettlemoyer, ``Llm. int8 (): 8-bit matrix multiplication for transformers at scale,'' \emph{arXiv preprint arXiv:2208.07339}, 2022.

\bibitem{hinton2015distilling}
G.~Hinton, O.~Vinyals, and J.~Dean, ``Distilling the knowledge in a neural network,'' \emph{arXiv preprint arXiv:1503.02531}, 2015.

\bibitem{aguilar2020knowledge}
G.~Aguilar, Y.~Ling, Y.~Zhang, B.~Yao, X.~Fan, and C.~Guo, ``Knowledge distillation from internal representations,'' in \emph{Proc. of the AAAI Conference on Artificial Intelligence}, vol.~34, no.~05, 2020, pp. 7350--7357.

\bibitem{lecun1989optimal}
Y.~LeCun, J.~Denker, and S.~Solla, ``Optimal brain damage,'' \emph{Advances in neural information processing systems}, vol.~2, 1989.

\bibitem{hoefler2021sparsity}
T.~Hoefler, D.~Alistarh, T.~Ben-Nun, N.~Dryden, and A.~Peste, ``Sparsity in deep learning: Pruning and growth for efficient inference and training in neural networks,'' \emph{The Journal of Machine Learning Research}, vol.~22, no.~1, pp. 10\,882--11\,005, 2021.

\bibitem{wang2021not}
Y.~Wang, R.~Huang, S.~Song, Z.~Huang, and G.~Huang, ``Not all images are worth 16x16 words: Dynamic transformers for efficient image recognition,'' \emph{Advances in Neural Information Processing Systems}, vol.~34, pp. 11\,960--11\,973, 2021.

\bibitem{avit}
L.~Meng, H.~Li, B.-C. Chen, S.~Lan, Z.~Wu, Y.-G. Jiang, and S.-N. Lim, ``Adavit: Adaptive vision transformers for efficient image recognition,'' in \emph{Proceedings of the IEEE/CVF Conference on Computer Vision and Pattern Recognition}, 2022, pp. 12\,309--12\,318.

\bibitem{wojcik2023adaptive}
B.~Wójcik, A.~Devoto, K.~Pustelnik, P.~Minervini, and S.~Scardapane, ``Adaptive computation modules: Granular conditional computation for efficient inference,'' 2023.

\bibitem{yang2022deep}
M.~Yang and H.-S. Kim, in \emph{ICASSP 2022-2022 IEEE International Conference on Acoustics, Speech and Signal Processing (ICASSP)}.\hskip 1em plus 0.5em minus 0.4em\relax IEEE, 2022, pp. 5193--5197.

\bibitem{dai2022nonlinear}
J.~Dai, S.~Wang, K.~Tan, Z.~Si, X.~Qin, K.~Niu, and P.~Zhang, ``Nonlinear transform source-channel coding for semantic communications,'' \emph{IEEE Journal on Selected Areas in Communications}, vol.~40, no.~8, pp. 2300--2316, 2022.

\bibitem{wang2023adaptive}
L.~Wang, W.~Wu, F.~Zhou, Z.~Yang, and Z.~Qin, ``Adaptive resource allocation for semantic communication networks,'' \emph{arXiv preprint arXiv:2312.01081}, 2023.

\bibitem{zhou2021semantic}
Q.~Zhou, R.~Li, Z.~Zhao, C.~Peng, and H.~Zhang, ``Semantic communication with adaptive universal transformer,'' \emph{IEEE Wireless Communications Letters}, vol.~11, no.~3, pp. 453--457, 2021.

\bibitem{deit}
H.~Touvron, M.~Cord, and H.~J{\'e}gou, ``Deit iii: Revenge of the vit,'' in \emph{European Conf. on Computer Vision}.\hskip 1em plus 0.5em minus 0.4em\relax Springer, 2022, pp. 516--533.

\bibitem{weight_cloning}
M.~Samragh, M.~Farajtabar, S.~Mehta, R.~Vemulapalli, F.~Faghri, D.~Naik, O.~Tuzel, and M.~Rastegari, ``Weight subcloning: direct initialization of transformers using larger pretrained ones,'' 2023.

\end{thebibliography}

\end{document}